\documentclass[aps,prx,superscriptaddress,twocolumn,tightenlines,showpacs,floatfix,amsmath,amssymb,pdftex]{revtex4-1}

\usepackage{preamble}

\begin{document}

\title{The topological invariants of rotationally symmetric crystals}
\author{Jans Henke}
\author{Mert Kurttutan}
\affiliation{Institute for Theoretical Physics Amsterdam and Delta Institute for Theoretical Physics,
University of Amsterdam, Science Park 904, 1098 XH Amsterdam, The Netherlands}
\author{Jorrit Kruthoff}
\affiliation{Stanford Institute for Theoretical Physics, Stanford University, Stanford, CA 94305}
\author{Jasper van Wezel}
\email{vanwezel@uva.nl}
\affiliation{Institute for Theoretical Physics Amsterdam and Delta Institute for Theoretical Physics,
University of Amsterdam, Science Park 904, 1098 XH Amsterdam, The Netherlands}
\date{\today}

\begin{abstract}
Recent formal classifications of crystalline topological insulators predict that the combination of time-reversal and rotational symmetry gives rise to topological invariants beyond the ones known for other lattice symmetries. Although the classification proves their existence, it does not indicate a way of calculating the values of those invariants. Here, we show that a specific set of concentric Wilson loops and line invariants yields the values of all topological invariants in two-dimensional systems with pure rotation symmetry in class AII. Examples of this analysis are given for specific models with two-fold and three-fold rotational symmetry. We find new invariants that relate to the presence of higher-order topology and corner charges.
\end{abstract}

\maketitle

{\bf Introduction} ---
In crystalline topological insulators, the presence or absence of symmetries allows for the emergence of a wide variety of topological phases. These are labelled by an equally wide variety of topological invariants, ranging from the Chern number~\cite{Thouless1982}, to the two-dimensional Fu-Kane-Mele (FKM) invariants~\cite{Kane2005a, Kane2005b, Yu2011, Alexandradinata2014}, the Lau-Brink-Ortix (LBO) or line invariants~\cite{LBO2016}, as well as invariant features of the Wilson loop spectrum describing higher-order and fragile topological insulators~\cite{Schindler_2018,BouhonSlager2019,Bradlyn2019, Kooi2019, Hwang2019, BouhonSlager2021, LangeSlager2021}. A unified, symmetry-based approach describing all of these topological phases was recently proposed~\cite{Kruthoff2016, J3_2019, Po2017, Po2017MSG, slager2013}. This shows that the FKM, LBO, higher order, and similar invariants may all be extracted from an algorithmic analysis of lattice symmetries and their effect on the structure of Berry curvature. Moreover, being a complete classification of all possible such invariants (as guaranteed by the underlying K-theory), it predicts that additional, as yet unidentified, invariants of the same type exist in various crystals, for example those with rotational symmetries in two dimensions~\cite{J3_2019}. Explicitly: in wallpaper groups $p2$, $p3$, $p4$, and $p6$ we expect to find 4, 2, 3, and 3 $\mathbb{Z}_2$ invariants, of which only 3, 1, 2 and 2 invariants are known.

Although the classification predicts the existence of topological invariants for systems with a given symmetry, it does not give a way of identifying or evaluating them in any specific system. Here, we introduce a single, unified diagnostic that yields the values of all curvature-based invariants in two-dimensional crystalline topological phases with time-reversal symmetry (TRS) and rotational symmetry, including the new phases predicted by the symmetry-based classification. Our analysis employs a spectrum of concentric Wilson loops rather than the usual spectrum of parallel Wilson loops~\cite{Yu2011, Alexandradinata2014,Alexandrinata2014_ref1, Bradlyn2019, Kooi2019}. The concentric loops are tailored to the rotational symmetry of the crystal lattice, allowing them to capture the full influence of symmetry on the topological structure.

We showcase the use of the concentric Wilson loop spectrum in examples with three-fold and two-fold rotational symmetry. We then indicate how the same methodology can be applied to any $n$-fold rotational symmetry group, and discuss the significance of the new invariant with respect to edge and corner states.

{\bf The concentric Wilson loop spectrum} ---
The gauge invariant eigenvalues of (non-Abelian) Wilson loops can be interpreted as generalising the (Abelian) Berry curvature to systems with internal and lattice symmetries~\cite{Schindler_2018}. We will briefly review the established use of Wilson loops for extracting the FKM invariant before introducing the set of concentric Wilson loops that we employ below to capture the full topology of rotationally invariant crystals.

In a material with $N$ occupied bands, the non-Abelian Berry connection is an $N \times N$ matrix with vector-valued components defined as $\mathbf{A}_{mn}(\mathbf{k}) = i \braket{u_m(\mathbf{k})|\mathbf{\nabla}_k|u_n(\mathbf{k})}$. Here, $\ket{u_n(\mathbf{k})}$ indicates an occupied Bloch state at momentum $\mathbf{k}$ with band index $n \in \{1,\dots, N\}$. The Wilson loop $\mathcal{W}[\mathcal{C}]$ on a closed contour $\mathcal{C}$ in the Brillouin zone (BZ) is then given by:
\be 
\mathcal{W}[\mathcal{C}] = \mathcal{P} \exp\left( i \oint_{\mathcal{C}} d\mathbf{k} \cdot \mathbf{A} \right),
\ee
with $\mathcal{P}$ indicating path ordering. The Wilson loop is a $U(N)$ matrix and satisfies $\mathcal{W}[\mathcal{C}]\mathcal{W}[\mathcal{C}]^{\dagger} = \mathcal{W}[\mathcal{C}]\mathcal{W}[\mathcal{C}_r] = 1$, with $\mathcal{C}_r$ the orientation-reversed loop. The Wilson loop can equivalently be expressed as a product of projectors onto the occupied states along the loop, which is particularly convenient when analysing its properties under symmetry transformations, as shown in Appendix~\ref{app:WL}. 

\begin{figure}[bt]
    \centering
    \includegraphics[width=\linewidth]{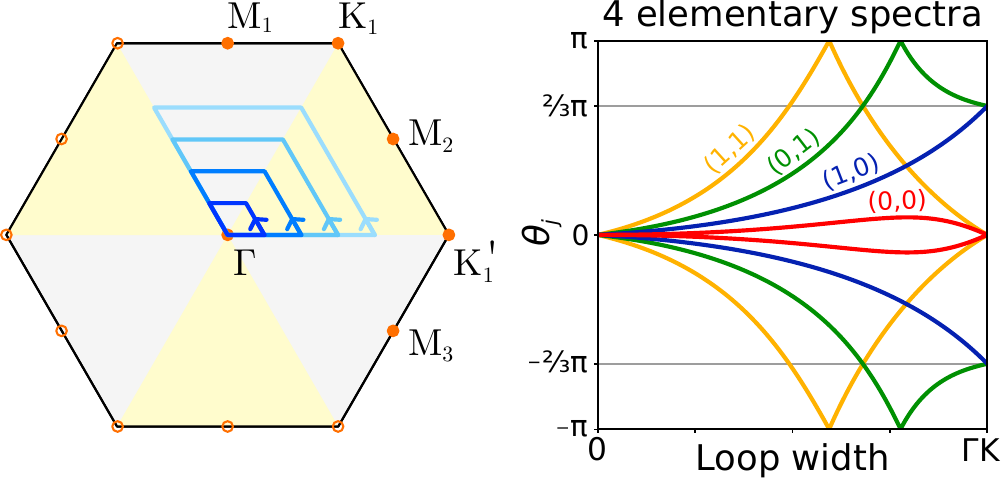}
    \caption{\label{fig:intro}\textit{Left:} Concentric Wilson loops in a three-fold rotationally symmetric BZ. \textit{Right:} The corresponding four elementary spectra (schematic), with topological indices indicated as $\mathbf{w}=(w_{\rm FKM}, w_\pi)$. Spectra with higher winding can be reduced to elementary ones upon addition of a topologically trivial Kramers pair with $\mathbf{w}=(0,0)$.}
\end{figure}

In systems without any internal or crystal symmetries, one can consistently assign band indices to all states, such that the off-diagonal elements of the Berry connection matrix are zero, and each band can be said to have its own independent Berry curvature and $U(1)$ Wilson loops. The presence of symmetries, however, may cause unavoidable degeneracies between bands, which necessitate the use of the non-Abelian Berry connection. Notably, TRS causes the formation of Kramers pairs, and the associated FKM invariant~\cite{Kane2005a, Kane2005b, FuKane2006} may be formulated in terms of a $U(2)$ Wilson loop~\cite{RyuLee2008}. In TRS systems without any additional symmetries that cause degeneracies of more than two bands (for instance due to non-symmorphic symmetries~\cite{Young2015}), it is always possible to consider the $U(2)$ Wilson loop, and more general Berry connections do not contain any additional topological information. We therefore restrict our attention to the $U(2)$ Wilson loops relevant for TRS crystals from here on.

The gauge-invariant Wilson loop eigenvalues of generic closed paths on the Brillouin torus are typically not quantized, and do not represent topological invariants~\footnote{The eigenvalues of any Wilson loop $\mathcal{W}[\mathcal{C}]$ are restricted to being a pure phase $\exp\left(i\theta_j\right)$, where $j$ denotes the eigenvalue index. In the text, we use the term ``eigenvalue'' to indicate the phase $\theta_j$.}. This may be different for paths respecting constraints imposed by symmetry. For instance, the partial polarisation in TRS systems is given by the (degenerate) $U(2)$ Wilson loop eigenvalues $\theta_j$ along a loop respecting the time-reversal symmetry~\cite{Alexandradinata2014}. Even without such symmetry constraints, however, one can gain topological information by considering a family of generic Wilson loop contours, parametrized by some variable $k$, that collectively yield a spectrum $\theta_j(k)$. For the family of contours in two dimensions with constant crystal moment component $k_x$, the parity of the winding of the Wilson loop spectrum (WLS) $\theta_j(k_x)$ equals the FKM invariant~\cite{Yu2011, Alexandradinata2014, Bradlyn2019}.

Here, we consider an alternative, `concentric' Wilson loop spectrum, which allows us to not only evaluate the generic FKM invariant, but also the newly-predicted invariant specific to crystals with rotational symmetry. The concentric spectrum consists of loops that do not cycle around the entire BZ torus, but instead grow from an infinitesimal loop to enclosing $(1/n)^{\rm{th}}$ of the $n$-fold rotationally symmetric BZ (see Fig.~\ref{fig:intro}). The gauge-invariant eigenvalues of each of the loops in the spectrum come in pairs of opposite sign and describe the net enclosed $U(2)$ Berry curvature. The spectrum trivially starts at zero (no curvature enclosed), and ends at a value corresponding to $\pm(1/n)^{\rm{th}}$ of the total $U(2)$ Berry curvature present in the BZ. The total winding of the spectrum, times $n$ and modulo $4\pi$, thus equals the FKM invariant. Additionally, we prove in Appendix~\ref{app:WL} that linear crossings at $\theta_j=\pi$ can only be gapped by pair-wise annihilation. The parity of the number of times the spectrum crosses $\theta_j=\pi$ therefore describes a second topological invariant.

Crossings through $\theta_j=0$ do not carry the same protection. To see this, consider a spectrum starting and ending at $\theta_j=0$ with one additional zero crossing in between. This spectrum may be reduced to a completely flat spectrum ($\theta_j=0$ for all loops in the spectrum) by transformations of the Hamiltonian that do not break any symmetries or constraints. In more general spectra the zero crossings are \emph{fragile}, in the sense that they can not generally be reduced to a completely flat spectrum by themselves, but the zero crossings can be removed (and the total winding reduced) after adding a trivial Kramers pair and letting the WLS hybridize. 

In fact, within each wallpaper group, one can identify a set of concentric WLS whose winding cannot be reduced any further upon the addition of topologically trivial, occupied pairs of bands. We call these elementary spectra. See Appendix~\ref{app:elementaryspectra} for details. Fig.~\ref{fig:intro} depicts the elementary spectra for the case of three-fold rotational symmetry. These elementary concentric WLS indicate both the FKM invariant, corresponding to the parity of the winding divided by $2\pi/3$ (or $2\pi/n$ for $n$-fold rotational symmetry) \footnote{To minimize numerical error, we consider a series of concentric hexagonal loops centered at $\mathbf{k}_\Gamma=(0,0)$ in explicit calculations. We then unfold the spectrum (to access eigenvalues of magnitude larger than $\pm\pi$), and divide the resulting values by three. This is equivalent to considering the spectrum of concentric loops on one third of the BZ, owing to the three-fold symmetry of the $U(2)$ Berry curvature.}, and the new invariant, corresponding to the parity of the number of $\pi$-crossings.

{\bf Three-fold rotation}\label{sec:p3} ---
To see how the concentric WLS is used to identify the value of the new invariant in practice, consider a crystal with three-fold rotational symmetry ($C_3$). In systems with $C_3$ and TRS, the Brillouin zone depicted in Fig.~\ref{fig:intro} hosts four time-reversal invariant momenta (TRIM), namely $\Gamma$ and $M_i$ (with $i=1,2,3$; mapped onto one another by $C_3$). The $K$-points are equivalent and symmetric under $C_3$, but mapped onto the $K'$-point under the action of the TRS operator $T$. In contrast to any even-fold rotational group, there is no combination of $C_3$ with $T$ that returns a state to the crystal momentum $\mathbf{k}$ it started with. Notice also that $T^2 = (C_3)^3 = -1$ because of the spin-one-half nature of electrons. The symmetry-based classification of crystalline topological insulators predicts two $\mathbb{Z}_2$ invariants for these three-fold symmetric systems ~\footnote{This corrects a statement in \cite{J3_2019}: although $U(1)$ vortices at $K$ cannot be moved, they can be smeared in a $C_3$ and TRS invariant fashion, leaving two rather than three $\mathbb{Z}_2$ invariants.}. 
Accordingly, we can describe all allowed topological phases in $C_3$-symmetric class AII systems by pairs of numbers $\mathbf{w} = (w_{{\rm FKM}},w_\pi)$.
The first of these corresponds to the FKM invariant, and the second is given by the parity of the number of $\pi$-crossings in this WLS. The latter has, to the best of our knowledge, not been identified before \footnote{Besides the new invariant not having been identified in any model systems on the basis of phenomenological observations, the complete and rigorous K-theory classification for TRS systems with three-fold rotational symmetry is, to the best of our knowledge, not known.}. 

As an example implementing these invariants, consider a TRS generalization of the Haldane model, given by the Bloch Hamiltonian~\cite{Haldane1988,Sticlet2013,Bhattacharya2017}:
\begin{align}
H &= \begin{pmatrix}
H_{\rm Hal}^+ & 0 \\
0 & H_{\rm Hal}^-
\end{pmatrix} \notag \\
H_{\rm Hal}^{\pm}(\mathbf{k}) &= d_1(\mathbf{k}) \tau_x + d_2(\mathbf{k}) \tau_y + d_3^{\pm}(\mathbf{k}) \tau_z.
\end{align}
Here, $\t_i$ are Pauli matrices for the sublattice degree of freedom, so that for example $\t_x=a_{\sigma}^\dagger b_{\sigma}+b_{\sigma}^\dagger a_{\sigma}$ with $a^\dagger_{\sigma}$ and $b^{\dagger}_{\sigma}$ creation operators for electrons with spin ${\sigma}$ on different sublattices. We also defined $d_1 = \sum_{j=1}^3 [t_1 \cos(\mathbf{k}\cdot\mathbf{a}_j) +t_3 \cos(\mathbf{k}\cdot\mathbf{c}_j)]$, $d_2 = \sum_{j=1}^3 [-t_1 \sin(\mathbf{k}\cdot\mathbf{a}_j) -t_3 \sin(\mathbf{k}\cdot\mathbf{c}_j)]$, and $d_3^{\pm} =  m \pm \sum_{j=1}^6 t_2(-1)^j \sin(\mathbf{k}\cdot\mathbf{b}_j)$. The vectors $\mathbf{a}_j$, $\mathbf{b}_j$ and $\mathbf{c}_j$ connect first, second, and third-nearest neighbours~\footnote{Explicitly, $\mathbf{a}_j = (Q_3^T)^j \{0,1\}^T$, $\mathbf{b}_j = (Q_6^T)^j \{\sqrt{3},0\}^T$ and $\mathbf{c}_j = (Q_3^T)^j \{0,-2\}^T$, where $Q_3$ and $Q_6$ are three and six-fold rotation matrices, respectively, and the lattice spacing is set to unity.}. The inclusion of hopping integrals up to third-nearest neighbours allows for phases of $H_{\rm Hal}$ with Chern numbers larger than one~\cite{Haldane1988,Sticlet2013,Bhattacharya2017}. To ensure the system is gapped, we add a Rashba-type spin-orbit coupling connecting the time reversed elements $H_{\rm Hal}^{\pm}$:
\be
    H_{\rm R} = i \lambda_{\rm{R}} \sum_{j=1}^3 \sum_{\sigma\sigma'} 
    (\mathbf{c}_j \times \mathbf{s})^{\sigma\sigma'}_z e^{i\mathbf{k}\cdot \mathbf{c}_j} a_{\sigma}^\dagger b_{\sigma'} + {\rm h.c.}
\ee
Here $\sigma$ is a spin index, $\mathbf{s}$ is the vector of Pauli matrices, and $a^\dagger$ and $b^{\dagger}$ are again creation operators for electrons on different sublattices. The extended Haldane model Hamiltonian, $H_{p3} = H + H_{\rm R}$, is invariant under TRS (${T} = i\sigma_y \otimes \tau_0\,\mathcal{K}$) and three-fold rotational symmetry (${C}_3 = \exp{\left({i\pi\sigma_z}/{3} \right)} \otimes \tau_0$).

\begin{figure}[t]
    \centering
    \includegraphics[width=\linewidth]{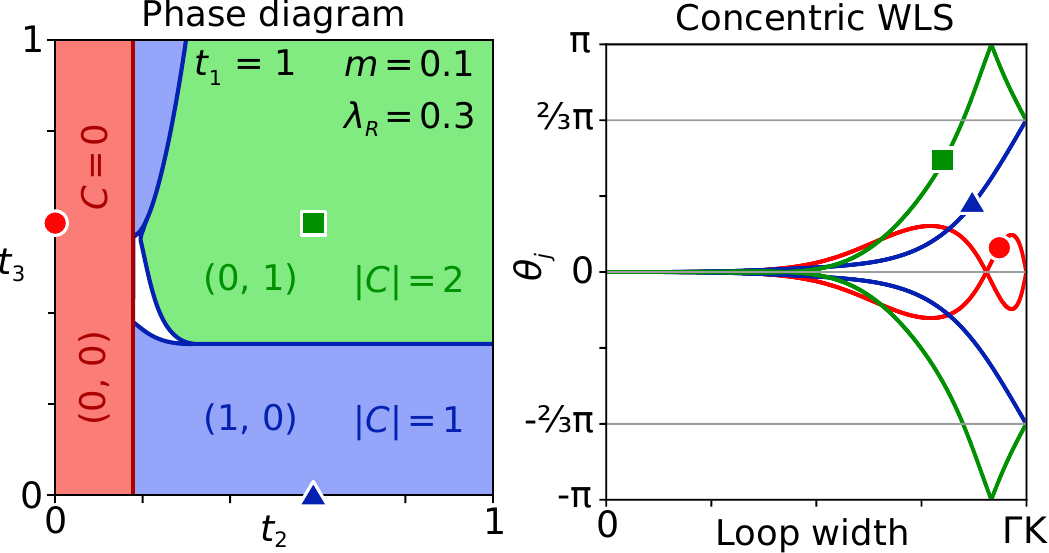}
    \caption{\label{fig:p3} \textit{Left}: The phase diagram of the extended Haldane model $H_{p3}$ with $t_1=1$, $m=0.1$, $\lambda_R=0.3$. The red phase has trivial values for all topological indices, i.e. $\mathbf{w} = 0$. The blue and green phases are non-trivial with their respective $\mathbf{w} = (w_{{\rm FKM}},w_\pi)$ indicated alongside the magnitude of the Chern numbers for $H_{\rm Hal}^{\pm}$ in each phase. In the white region the phase could not be unambiguously determined due to (the proximity to) a gap closing. \textit{Right}: The concentric WLS over one third of the BZ for the three points indicated by corresponding symbols in the phase diagram. The values of the two topological invariants for these phases correspond to the parity of $3/(2\pi)$ times the winding of the WLS value, and the parity of the number of $\pi$ crossings in the spectrum.}
\end{figure}

As shown in Fig. \ref{fig:p3}, the extended Haldane model hosts three distinct topological phases, which can be accessed by varying the relative magnitude of the hopping parameters $t_n$. The right-hand side of Fig. \ref{fig:p3} shows an example spectrum for each phase. The red phase is trivial, having zero total winding and zero $\pi$-crossings. The blue and green phases are non-trivial and have a total winding that is an odd multiple of $2\pi/3$ and an odd number of $\pi$-crossings, respectively. Their topological indices are thus $\mathbf{w} = (1,0)$ and $\mathbf{w} = (0,1)$. Also indicated are the Chern numbers of the individual $H_{\rm Hal}^{\pm}$ Hamiltonians, which equal $w_{{\rm FKM}}$ modulo 2. 

{\bf Two-fold rotation}\label{sec:p2} ---
We now apply the same concentric WLS approach to a TRS system with two-fold rotational symmetry. The topological classification for this class of systems was predicted to be $\mathbb{Z}_2^4$~\cite{J3_2019, Kruthoff2016}. The K-theory underlying this classification~\cite{Freed:2012uu} indicates that two of the $\mathbb{Z}_2$ invariants are \emph{strong} in the K-theory sense of not requiring translational symmetry, while the remaining two invariants do depend on the presence of a periodic lattice~\cite{Stehouwer2018} (see also~\cite{Cornfeld_2021, Shiozaki:2018srz} for related K-theoretic calculations). These latter two may be interpreted as LBO line invariants along the lines $k_x=0$ and $k_y=0$~\cite{LBO2016}, while one of the strong invariants coincides with the usual FKM invariant. In $p2$, it has recently been shown that a fourth invariant can be identified in systems with trivial FKM invariant~\cite{Kooi2019}, which we identify with our new invariant $w_\pi$ (see also Appendix~\ref{app:p2}). These four invariants are independent of one another, and cannot be changed upon the addition of a trivial band. Topological phases in $C_2$-symmetric class AII systems can thus be labelled by $\mathbf{w}=(w_{\rm FKM}, w_\pi, w_{\rm LBO_x}, w_{\rm LBO_y})$.

\begin{figure}[bt]
    \centering
    \includegraphics[width=0.95\linewidth]{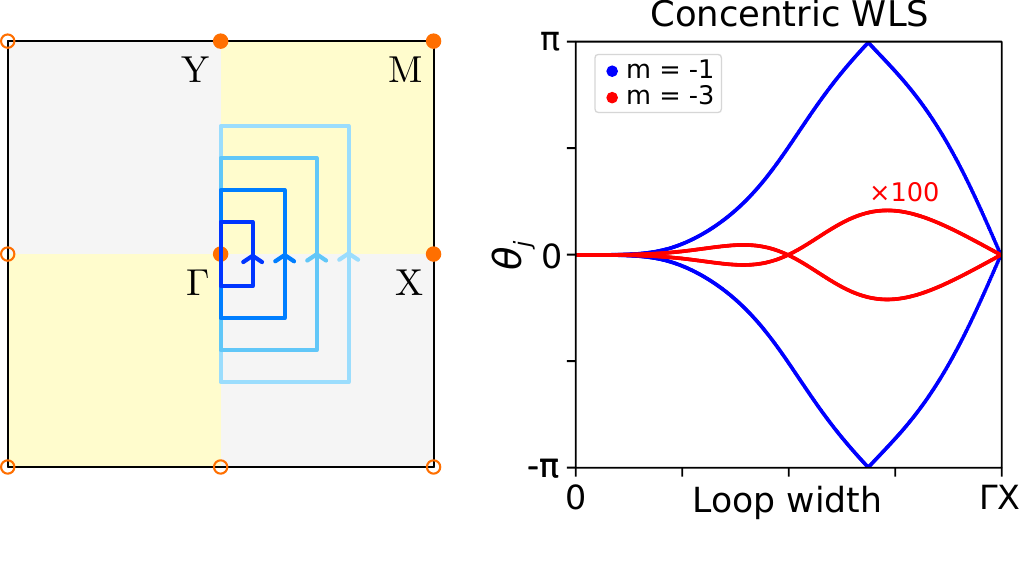}
    \vspace{-7pt}
    \caption{\label{fig:p2}\textit{Left:} Concentric Wilson loops covering half the two-fold rotationally symmetric BZ. \textit{Right:} The concentric Wilson loop spectrum of $H_{p2}$, with $t_1=1/2$, $t_2=1/10$, and two different values of parameter $m$. The two different topological phases can be described by $\mathbf{w}=(w_{\rm FKM}, w_\pi, w_{\rm LBO_x}, w_{\rm LBO_y})$; red has $\mathbf{w}=(0,0,0,0)$, and blue has $\mathbf{w}=(0,1,0,0)$.}
\end{figure}

Just as in the case of $p3$, the concentric WLS of systems in $p2$ may always be reduced to four elementary spectra upon addition of trivial occupied Kramers pairs. The complex conjugate eigenvalues of the concentric Wilson loops make up a WLS that always starts at $\theta_j = 0$, and ends at either $\theta_j = 0$ or $\pm\pi$, corresponding respectively to $w_{\rm FKM}=0$ or 1. As before, zero-crossings may be removed (possibly upon addition of a trivial Kramers pair) by flattening the entire spectrum, while $\pi$-crossings are topologically protected (see Appendix~\ref{app:WL}). Notice that this is true even for $\pi$-crossings in a spectrum of the green type in Fig.~\ref{fig:intro} (shown for $p2$ in Appendix~\ref{app:elementaryspectra}), which will have a $\pi$-crossing followed by the spectrum ending at $\theta_j=\pi$. Although the $\pi$-crossing can be pushed towards the end of the spectrum, it cannot be annihilated and gapped there, because the value at the end of the spectrum equals the FKM invariant, and is quantised. Therefore, the parity of $\pi$-crossings constitutes a topological invariant even in this special case.

\begin{figure*}[t]
    \centering
    \includegraphics[width=0.8\linewidth]{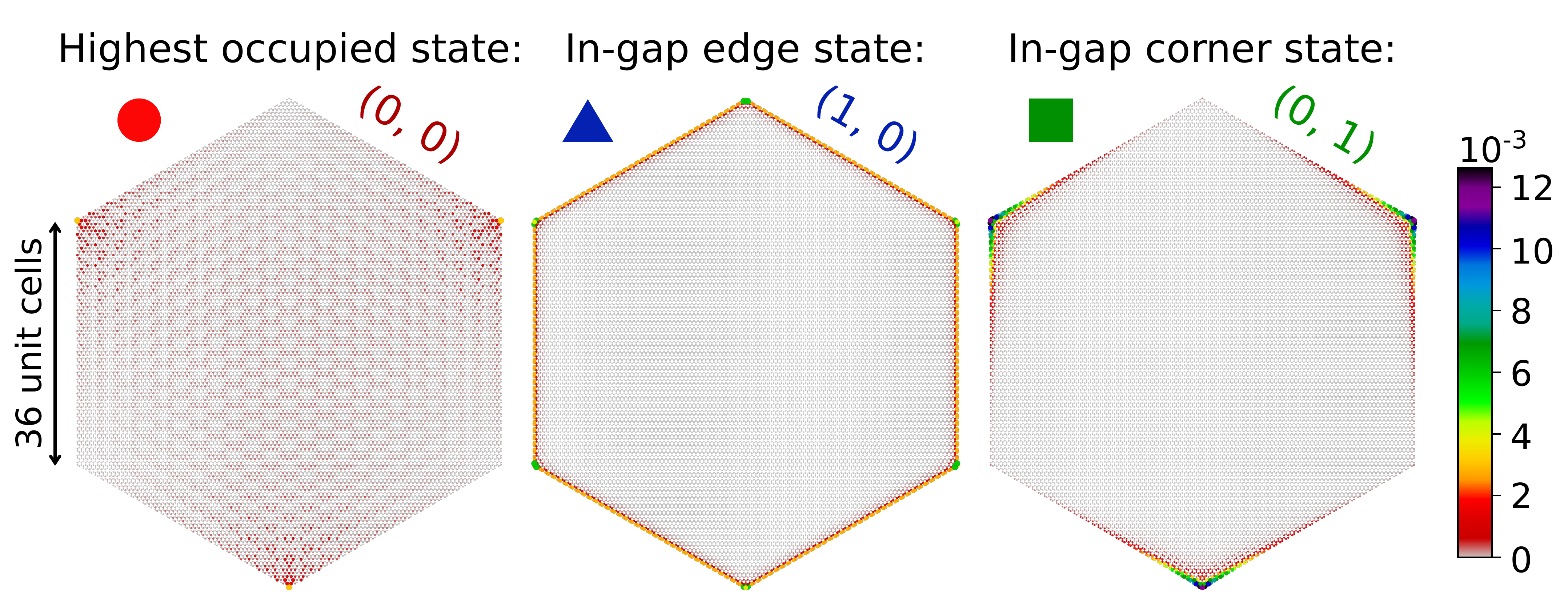}
    \caption{\label{fig:finite} Exemplary states of the extended Haldane model in a finite-sized hexagon with armchair edges, for the three topological phases with parameter values as indicated by the symbols in Fig. \ref{fig:p3}. The (0,0) phase hosts no in-gap states, while the (1,0) phase hosts in-gap edge states and the (0,1) phase hosts in-gap corner states. The colour scale indicates the local density of states, normalised to one in each panel. The states shown lie at $E/t_1=-0.2$, $0$, and $-0.05$, respectively.}
\end{figure*}

As a generic example of the application of concentric WLS to identify all invariants in two-fold symmetric systems, we consider a TRS version of the Qi-Wu-Zhang model~\cite{QiWuZhang2006}. We define the Bloch Hamiltonian:
\begin{align}
    H_{\rm QWZ}(\mathbf{k}) &= d_1\, \sigma_0 \otimes \tau_x + d_2\, \sigma_z \otimes \tau_y + d_3\, \sigma_0\otimes \tau_z.
\end{align}
Here, $\sigma_i$ and $\tau_i$ are Pauli matrices associated with spin and sublattice respectively. The coefficients are given by $d_1(\mathbf{k}) = \cos k_x$, $d_2(\mathbf{k}) = \cos k_y$, and $d_3(\mathbf{k}) = m^2 - (\sin k_x + \sin k_y)^2$. This Hamiltonian has two chiral symmetries, ${C}_{x,y} = \sigma_{x,y} \otimes \tau_y$, which we break by adding:
\begin{equation}
    H_{\rm int}(\mathbf{k}) = t_1 \sin k_x\, \sigma_x \otimes \tau_z + t_2 \sin(k_x + k_y)\, \sigma_y \otimes \tau_z
\end{equation}
We refer to the full model $H_{p2}=H_{\rm QWZ} + H_{\rm int}$ as the QWZ model. It has time-reversal symmetry (${T} = i\sigma_y \otimes \tau_0\,\mathcal{K}$, with $\mathcal{K}$ complex conjugation) and two-fold rotational symmetry (${C}_2 = i\sigma_z \otimes \tau_0$). In the following we consider parameter values $t_1 = 1/2$ and $t_2 = 1/10$.  

As shown in Fig.~\ref{fig:p2}, the concentric WLS of loops covering half of the BZ indicates that depending on the value of parameter $m$, the QWZ model resides in one of two phases, with the newly identified invariant $w_\pi = 0$ or 1. In both phases $w_{\rm FKM} = 0$, while direct computation of the LBO line invariants, equivalent to the value of the (degenerate) WL eigenvalues of loops cycling around the BZ torus with $k_x=0$ or $k_y=0$, shows that these are also zero in both phases.

{\bf Edge states and corner charges}\label{sec:edgecorner} ---
The bulk-boundary correspondence suggests edge states to be present in finite-sized materials whose bulk Hamiltonian has a non-trivial FKM invariant. Recently, there has been a proposal for a bulk-corner correspondence in rotationally-symmetric topological insulators~\cite{Kooi2021bulkcorner}. Intuitively, one expects a minimal requirement for topological corner charges to appear to be the presence of a winding WLS -- indicating the presence of a non-trivial topology -- with a trivial FKM invariant such that there are no protected edge states. Such a situation corresponds precisely to the $(w_{\rm FKM}, w_\pi) = (0,1)$ phase as seen in the models considered here. To see if this phase indeed hosts corner charges, we consider the extended Haldane model both on a finite-sized hexagon (Fig. \ref{fig:finite}) and in ribbon configurations with either armchair or zigzag type edges (see Appendix \ref{app:ribbon})~\cite{Pybinding}, for the same sets of parameter values as highlighted by symbols in Fig. \ref{fig:p3}.

As expected, the $\mathbf{w}=(0,0)$ phase is a trivial insulator, and exhibits no edge or corner states in any configuration. The $(1,0)$ phase hosts topological edge states along both types of edges, as expected for a non-trivial FKM invariant. Interestingly, infinitely long ribbons with zigzag edges in the $(0,1)$ phase also host edge states, while the edge states along ribbons with armchair edges become gapped. The finite-sized hexagon with armchair edges shown in Fig. \ref{fig:finite} exhibits fully localised corner charges, whose energies lie between those of the gapped edge states. 

As discussed in Appendix \ref{app:ribbon}, TRS systems derived from subsystems with odd Chern numbers ($w_{{\rm FKM}}=1$) will always host at least one pair of gapless edge states. Those derived from subsystems with even (non-zero) Chern numbers ($w_{\rm FKM}=0$) on the other hand, have no protection against their edge states being gapped. Whether the edge states are in fact gapped depends on the specific geometry of the finite-sized system considered. Even if the system has a gap in both the bulk and edge state spectrum, the existence of corner charges additionally requires their energies to lie within that gap. Depending on the parameters chosen, corner charges may be pushed into the gapped edge state spectrum and hybridise. A non-trivial value of the newly identified invariant ($w_\pi = 1$) thus signals the possibility of corner charges emerging, but whether they are realised in any specific finite-sized system depends on its detailed configuration.

{\bf Discussion} ---
The concentric Wilson loop spectrum serves as a single, unified diagnostic allowing the simultaneous evaluation of both the well-known FKM invariant, and the topological invariant that was predicted to exist in rotationally symmetric systems based on K-theoretical arguments~\cite{J3_2019, Stehouwer2018}. Complementing the concentric Wilson loop spectrum with LBO line invariants, which are given by the eigenvalues of high-symmetry Wilson loops, yields the values of all possible topological invariants appearing in the complete classification of TRS systems with rotational symmetry~\cite{J3_2019}. 

We gave explicit examples of this construction for particular tight-binding models in class AII with two-fold and three-fold rotational symmetries, but stress that the approach is general and can also be applied to cases with higher rotational symmetries, as detailed in Appendix~\ref{app:elementaryspectra}. Combining rotations with additional symmetries, including non-symmorphic ones, may be expected to further enrich the analysis.

In wallpaper group $p3$, we showed that an odd number of $\pi$-crossings in the concentric Wilson loop spectrum signals the possibility of localised corner charges appearing in finite-sized samples. This may explain the recent observation of corner charges in specific rotationally symmetric systems, which were suggested to be related to a form of fragile topology~\cite{Kooi2021bulkcorner}. Because the parity of $\pi$-crossings in the concentric Wilson loop spectrum is unaffected by the addition of topologically trivial Kramers pairs, the present analysis in fact suggests the bulk-corner correspondence in these systems to be stable and described by a true, rather than fragile, topological invariant. While further investigation is required, the existence of this stable invariant is not visible in the usual linear Wilson loop spectrum, underlining the utility of the concentric Wilson loops.

\bibliography{bibliography}

\appendix
\counterwithin{figure}{section}
\onecolumngrid

\section{Topological invariants in the concentric Wilson loop spectrum}\label{app:WL}

Following the arguments in Refs.~\cite{Yu2011, BouhonSlager2019, Kooi2019, Bradlyn2019}, we define the Wilson loop operator as a matrix-valued one-form for a given path $\mathcal{C}$ and Berry connection $\mathbf{A}(\mathbf{k})$:
\be 
\mathcal{W}(\mathbf{k}_0,\mathcal{C}) = \mathcal{P} \exp\left(i\oint_\mathcal{C} d\mathbf{k} \cdot \mathbf{A}(\mathbf{k})\right).
\ee
Here, $\mathcal{P}$ denotes path-ordering, $\mathbf{k}_0$ is a base point of $\mathcal{C}$, and the Berry connection has matrix elements $\mathbf{A}_{mn}(\mathbf{k}) = i \braket{u_m(\mathbf{k})|\mathbf{\nabla}_k|u_n(\mathbf{k})}$. It can be readily checked that the Berry connection is a skew-Hermitian matrix, rendering the Wilson loop a unitary operator. 

Under a unitary symmetry $g \in G$ the Wilson loop matrix elements transform as:
\be 
g: \mathcal{W}_{ij}(\mathbf{k}_0,\mathcal{C}) \to D_i^l(g,\mathbf{k}_0)\mathcal{W}_{lk}(g\cdot \mathbf{k}_0,g\cdot \mathcal{C})D^k_j(g,\mathbf{k}_0)^*.
\label{Wilson_sym}
\ee
Here $D^i_l(g,\mathbf{k}_0)$ is the representation of the symmetry $g$, which could depend on base point $\mathbf{k}_0$. If $\mathcal{C}$ is a closed loop, the two $D$'s in this expression have the same $\mathbf{k}_0$ argument, while for a line it would be the the beginning and end point, respectively on the right and left. Invariance under a unitary symmetry thus means
\be 
\mathcal{W}_{ij}(\mathbf{k}_0,\mathcal{C}) = D_i^l(g,\mathbf{k}_0)\mathcal{W}_{lk}(g\cdot \mathbf{k}_0,g\cdot \mathcal{C})D^k_j(g,\mathbf{k}_0)^*.
\ee
In the presence of TRS and additional symmetries, two types of choices for $\mathcal{C}$ are noteworthy. These are non-contractible cycles around the BZ (the type most commonly considered in the literature~\cite{Yu2011, Alexandradinata2014, Bradlyn2019, Kooi2019}) and loops that enclose high symmetry points (see also~\cite{Bradlyn2019}). The latter type can be chosen to be invariant under time-reversal symmetry. The loops around high-symmetry points can be considered non-contractible, because the high-symmetry points are fixed points of the spatial symmetries. 

To see how TRS affects the Wilson loop operators along these two types of non-contractible loops, we first discretize the Wilson loop, implementing the path-ordering explicitly. The curve $\mathcal{C}$ specifies a path in the BZ that can be parametrized by $\mathbf{k}(t)$ with $0\leq t \leq 1$ and $\mathbf{k}(0) = \mathbf{k}_0 = \mathbf{k}(1)$ (up to a reciprocal lattice vector). Conventionally, we will take $\mathbf{k}(t)$ decreasing with $t$, so we go around $\mathcal{C}$ in a clockwise direction. We discretize the curve by considering discrete values $t_i=i/N$ with $i = 1,2, \dots N$ and $N$ large. Denoting $\mathbf{k}_i$ the momentum at $t = t_i$, we then have:
\begin{align}
    \mathcal{W}_{mn}(\mathbf{k}_0,\mathcal{C}) &= \left[ \mathcal{P} \exp\left(i\oint_\mathcal{C} d\mathbf{k} \cdot \mathbf{A}(\mathbf{k})\right) \right]_{mn} \nonumber \\ 
    &\approx \sum_{j_1,j_2,\dots j_N}(\delta_{mj_1} + i\,d\mathbf{k}_1 \cdot \mathbf{A}_{mj_1}(\mathbf{k}_0))(\delta_{j_1 j_2} + i\,d\mathbf{k}_2 \cdot \mathbf{A}_{j_1 j_2}(\mathbf{k}_1)) \cdots (\delta_{j_N n} + i\,d\mathbf{k}_N \cdot \mathbf{A}_{j_N n}(\mathbf{k}_{N-1})).
\end{align}
Here we used $d\mathbf{k}_i \equiv \mathbf{k}_{i-1}-\mathbf{k}_{i}$. Using the definition of the Berry connection, we can write $\delta_{mn}- i\,(\mathbf{k}_i-\mathbf{k}_{i-1}) \cdot \mathbf{A}_{mn}(\mathbf{k}_i) \approx \braket{u_m(\mathbf{k}_{i-1})|u_n(\mathbf{k}_i)}$.
Taking $N \to \infty$ and keeping only terms linear in $d\mathbf{k}_i$, we then find:
\be 
    \mathcal{W}_{mn}(\mathbf{k}_0,\mathcal{C}) = \lim_{N\to \infty} \braket{u_m(\mathbf{k}_0)|\prod_{i=1}^N P(\mathbf{k}_i)|u_n(\mathbf{k}_0)}.
\ee
Here, the product is path-ordered around $\mathcal{C}$ and the projectors are defined as  $P(\mathbf{k}_i) = \sum_{j} \ket{u_j(\mathbf{k}_i)}\bra{u_j(\mathbf{k}_i)}$. 

To find the action of time-reversal symmetry on this representation of the Wilson loop operator, recall that time-reversal maps states at $\mathbf{k}$ onto states at $-\mathbf{k}$. We may therefore write a Bloch state at $-\mathbf{k}$ in terms of time-reversed Bloch states at $\mathbf{k}$:
\be 
    \ket{u_n(-\mathbf{k})} = \sum_m D_{nm}(\mathbf{k})T\ket{u_m(\mathbf{k})}
\ee
Here, the time reversal operator $T$ includes complex conjugation and $D$ is a unitary matrix. Using this relation, the projectors $P$ at $-\mathbf{k}$ can also be related to ones at $\mathbf{k}$:
\begin{align}
    P(-\mathbf{k}) &= \sum_{j,m,m'}D_{jm}(\mathbf{k})T\ket{u_m(\mathbf{k})}\bra{u_{m'}(\mathbf{k})}T^{\dagger}(D_{m'j})^*(\mathbf{k})\nonumber\\
    &= -\sum_m T\ket{u_m(\mathbf{k})}\bra{u_m(\mathbf{k})}T^{-1}\nonumber\\
    &= -TP(\mathbf{k})T^{-1}.
\end{align}
Here we used the fact that $T$ is anti-unitary, such that $T^{\dagger} =- T^{-1}$. If we now consider a loop $\mathcal{C}$ that is time-reversal symmetric, then:
\be 
    T\mathcal{W}(\mathbf{k}_0,\mathcal{C})T^{-1} = \mathcal{W}(-\mathbf{k}_0,\mathcal{C}),
\ee
To see this explicitly, take for example four points $\mathbf{k}_j$, $j=0,1,2,3$, on the unit circle in $\mathbf{k}$-space, equally spaced. Under $T$ we then have $\mathbf{k}_j \to \mathbf{k}_{j+2}$, so the discrete Wilson loop $\mathcal{W} = P(\mathbf{k}_0)P(\mathbf{k}_1)P(\mathbf{k}_2)P(\mathbf{k}_3)$
gets mapped to: 
\be 
    T\mathcal{W}T^{-1} = TP(\mathbf{k}_0)T^{-1}TP(\mathbf{k}_1)T^{-1}TP(\mathbf{k}_2)T^{-1}TP(\mathbf{k}_3)T^{-1} = P(\mathbf{k}_2)P(\mathbf{k}_3)P(\mathbf{k}_0)P(\mathbf{k}_1).
\ee
This is the same discrete Wilson loop, with the four points rotated but their order remaining the same. 

For a rotation symmetry $C_n$ we analogously find:
\be 
C_n\mathcal{W}(\mathbf{k}_0,\mathcal{C})C_n^{-1} = \mathcal{W}(g\cdot \mathbf{k}_0,g\cdot \mathcal{C}),
\ee
in agreement with the expression in Eq.~\ref{Wilson_sym}.
If we consider a loop for which the base-point dependence is only through its length $|\mathbf{k}_0|$, the minus sign in $-\mathbf{k}_0$ is irrelevant. This is the case for two-dimensional loops that are rotationally symmetric. 

To relate these results to previous literature, notice that often loops around non-contractible cycles are considered, and that those loops are in general neither invariant under rotations, nor time-reversal symmetric. In particular for the action of time-reversal symmetry, this means that for such loops we have:
\be 
T\mathcal{W}(\mathbf{k}_0,\mathcal{C})T^{-1} = \mathcal{W}(-\mathbf{k}_0,\mathcal{C}_r) = \mathcal{W}(-\mathbf{k}_0,\mathcal{C})^{\dagger}.
\ee
Here $\mathcal{C}_r$ is the orientation-reversed loop. That the orientation is reversed under time-reversal can be easily seen by considering for example a curve at constant $k_x$.

Now let us consider systems with both $C_3$ and $T$ symmetry, and consider a path $\mathcal{C}$ tracing out a hexagon centered at $\G$ with size $|\mathbf{k}_0|$. We denote the loop by $\mathcal{C}_h$. Imposing invariance under time-reversal and three-fold rotation, we then have the constraints:
\be
T\mathcal{W}(|\mathbf{k}_0|,\mathcal{C}_h)T^{-1} = \mathcal{W}(|\mathbf{k}_0|,\mathcal{C}_h),\quad C_3\mathcal{W}(|\mathbf{k}_0|,\mathcal{C}_h)C_3^{-1} = \mathcal{W}(|\mathbf{k}_0|,\mathcal{C}_h).
\ee
We can choose a gauge in which $T = i\sigma_y K$ and $C_3 = e^{\sigma_z i\pi/3}$, and consider the $U(2)$ Wilson loop of a single Kramers pair. The unitary matrix $\mathcal{W}$ can then be written as:
\be
\mathcal{W}_h(\mathbf{k}_0) = \mathcal{W}(|\mathbf{k}_0|,\mathcal{C}_h) = \begin{pmatrix}
a(\mathbf{k}_0)& b(\mathbf{k}_0)\\
c(\mathbf{k}_0) & d(\mathbf{k}_0).
\end{pmatrix}
\ee
The symmetry constraints on the matrix elements then become:
\be 
a(\mathbf{k}_0) = d^*(-\mathbf{k}_0),\quad b(\mathbf{k}_0) = -c^*(-\mathbf{k}_0),\quad a(\mathbf{k}_0) = a(g\cdot \mathbf{k}_0),\quad b(\mathbf{k}_0) = b(g\cdot \mathbf{k}_0) e^{-2\pi i/3}
\ee 
Here, $g$ is the action of the three-fold rotation symmetry on $k$-space. The first two constraints come from time-reversal symmetry, while the latter two arise from the three-fold rotational symmetry. The Wilson loop only depends on the length of $\mathbf{k}_0$, and since $|g\cdot \mathbf{k}_0| = |\mathbf{k}_0|$, $b$ must vanish in our chosen gauge. This means $c$ also vanishes and $d = a^*$. Furthermore, since the matrix $\mathcal{W}(\mathbf{k}_0)$ must be unitary, the $a(|\mathbf{k}_0|)$ must be a pure phase $e^{i\theta(|\mathbf{k}_0|)}$. The Wilson loop operator thus reduces to: 
\be \label{Wfinal}
\mathcal{W}_h(\mathbf{k}_0) = \mathcal{W}_h(|\mathbf{k}_0|) = \begin{pmatrix}
e^{i\theta(|\mathbf{k}_0|)} & 0 \\
0 & e^{-i\theta(|\mathbf{k}_0|)}
\end{pmatrix} = e^{i\sigma_z \theta(|\mathbf{k}_0|)}.
\ee
Because the Wilson loop operator is the exponent of only a single Pauli matrix, its eigenvalues are always complex conjugates and linear crossings of eigenvalues $\theta(|k\approx k_*|) = a + b (k-k_*) + \dots$ with $a \in \{0,\pm \pi\}$ are protected. This means that they can be gapped only by adding additional Pauli matrices that break either the $T$ or $C_3$ symmetry, or by having the coefficient $b$ vanish. The latter mechanism requires either two linear crossings to merge and annihilate, or a complete vanishing of the $k$-dependence of $\theta(k)$.

\begin{figure}[t]
    \centering
    \includegraphics[width=0.8\linewidth]{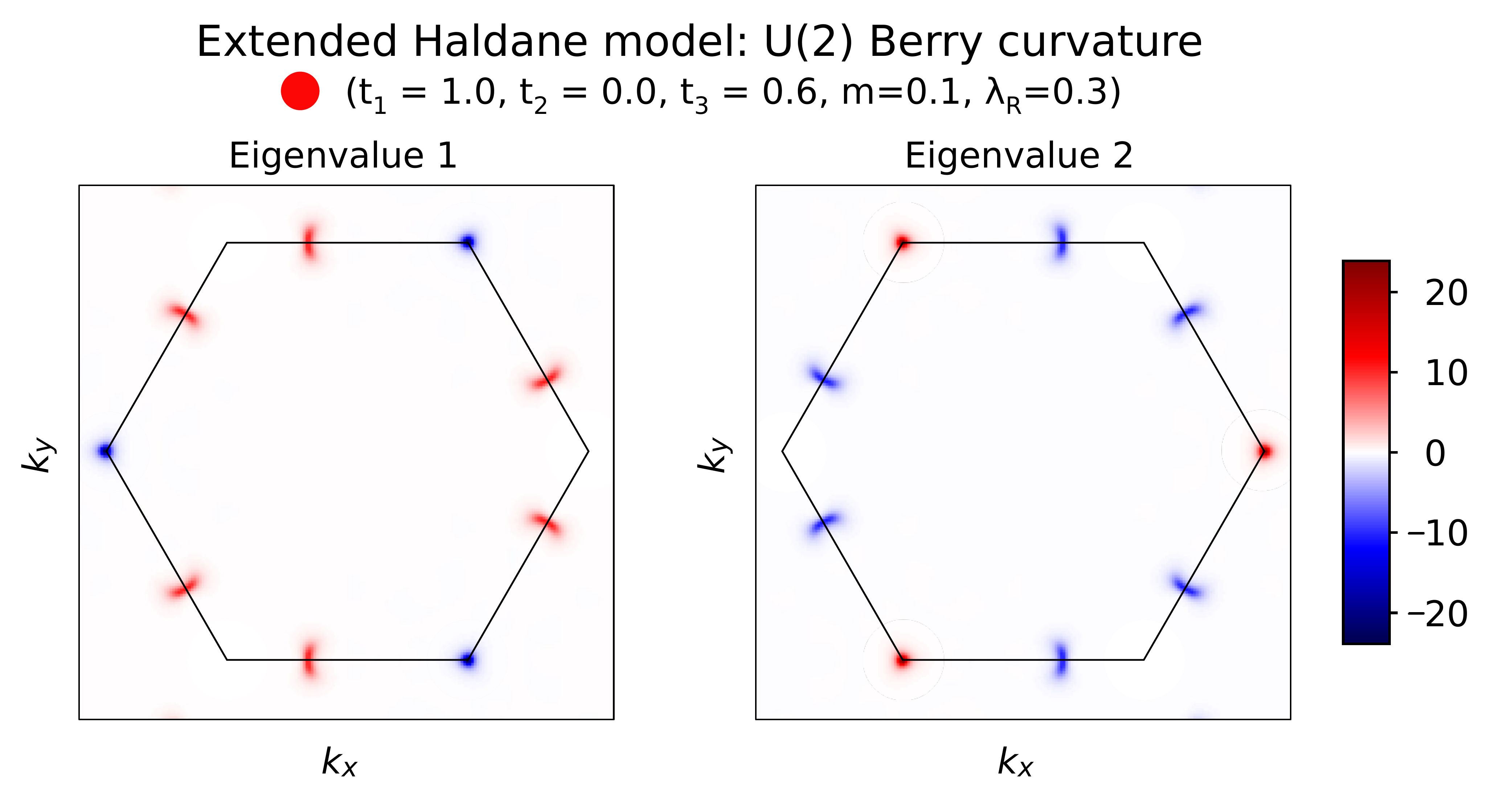}
    \caption{\label{fig:Wcurv} $U(2)$ Berry curvature of the extended Haldane model (see main text). The black hexagons indicate the size of the BZ. The parameters chosen here correspond to the red circle in Figure \ref{fig:p3}a. The curvature was computed as a grid of $200\times200$ square Wilson loops covering the area shown above.}
\end{figure}

Importantly, a full hexagonal Wilson loop contains redundant information due of the symmetries of the system, such that not all protected degeneracies in its spectrum are relevant. To extract meaningful topological information from a concentric WLS based on these types of loops, one must consider the concentric WLS of a single fundamental domain of the BZ, as defined by the symmetries. 
To determine what constitutes a fundamental domain, it is helpful to consider an infinitesimal Wilson loop, i.e. the $U(2)$ Berry curvature $\mathcal{F}_W$, at an arbitrary point $\mathbf{k}$ in the BZ. Note that the eigenvalues of such a generic infinitesimal loop are not constrained to be complex conjugates. Threefold rotational symmetry ensures that $\mathcal{F}_W (\mathbf{k}) = \mathcal{F}_W (g \cdot\mathbf{k})$, while TRS ensures that $\mathcal{F}_W (\mathbf{k}) = \mathcal{F}_W (-\mathbf{k})^\dagger$. That is, the curvature eigenvalues of either of the two states in the Kramers pair considered are three-fold symmetric, and the curvature of one of the two bands is inverted with respect to  the other (see Figure \ref{fig:Wcurv}). 
It then follows that neighbouring sextants of the BZ (may) have different curvatures, requiring the combination of two sextants of the BZ to capture all the symmetry-allowed information. From this we conclude that the fundamental domain of a $p3$ system covers two inequivalent sextants, and the relevant topological information can be extracted from one third of the full hexagonal WLS. More generally, for a system with (only) $n$-fold rotational symmetry, the fundamental domain constitutes 1/$n^{\rm th}$ of the BZ. 

Finally, one should note that a trivial Wilson loop spectrum, i.e. one starting and ending at zero, without crossing $\theta=\pi$ in between, is allowed by symmetry to flatten entirely, thus removing any crossings at $\theta=0$. As noted above, this corresponds to tuning the prefactor $b$ of the linear term in an expansion around the crossing point to zero. This complete flattening is only allowed for trivial spectra, but zero-crossings in any spectrum may be removed upon addition of a trivial Kramers pair, as explained in Appendix~\ref{app:elementaryspectra}. In that sense, zero-crossings are ``fragile''~\cite{Po2018fragiletopology, Bradlyn2019, Hwang2019, Kooi2019}. Because the starting point of a concentric spectrum is fixed to $\theta=0$, $\pi$-crossings do not have the same fragility, and can only be removed by either breaking a symmetry or closing the spectral gap. Thus, we conclude that both the winding (FKM invariant) and the parity of $\pi$-crossings in a concentric WLS covering a fundamental domain of the BZ, constitute topological invariants. 

\section{Multiple occupied bands and elementary spectra}\label{app:elementaryspectra}

Having found that the parity of crossings of eigenvalues at $\theta=\pi$ in a concentric WLS cannot be changed without breaking either TRS or rotational symmetry, we must ensure that this feature is also robust upon the addition of topologically trivial TRS-related pairs of bands (Kramers pairs). It is important to note here that when there are multiple occupied Kramers pairs, in the presence of only time-reversal and pure rotational symmetries, one can always consider the $U(2)$ WLS of each separate Kramers pair. If two Kramers pairs overlap, such that the identification of band indices is non-trivial, it is always possible to perturb the system without breaking any symmetries and separate the overlapping pairs. Each individual $U(2)$ WLS is subject to the symmetry restrictions discussed in Appendix~\ref{app:WL}, and can be assigned its own topological invariants.

\begin{figure}[t]
    \centering
    \includegraphics[width=0.75\linewidth]{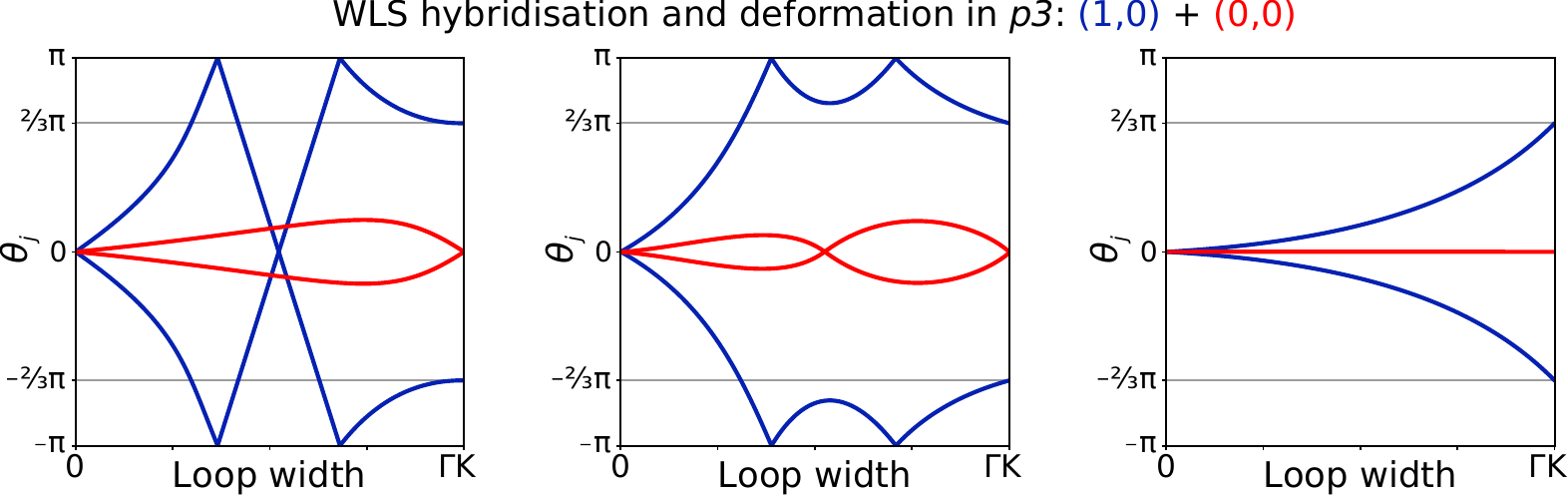}
    \caption{\label{fig:morebands} Sketch of an allowed hybridisation and deformation process of a WLS with $(w_{\rm FKM}, w_\pi) = (1,0)$ (blue) in the presence of a second, trivial $(0,0)$ (red) occupied Kramers pair. Importantly, the topological class remains the same throughout.}
\end{figure}

Adiabatic deformations of the WLS of an individual Kramers pair will not change the topological indices associated with that spectrum. However, some (quantised) amount of $U(2)$ Berry curvature in one Kramers pair may transfer to another occupied pair as crossings between two $U(2)$ WLS open up a gap. Different combinations of spectra for multiple occupied Kramers pairs may therefore represent the same combined topological phase (see Fig.~\ref{fig:morebands}). By simply summing the topological invariants extracted from each occupied Kramers pair's concentric WLS and (where applicable) line invariants, one can obtain the total topological classification of any Hamiltonian with $T$ and $C_n$ symmetries. 

\begin{figure}[ht]
    \centering
    \includegraphics[width=0.55\linewidth]{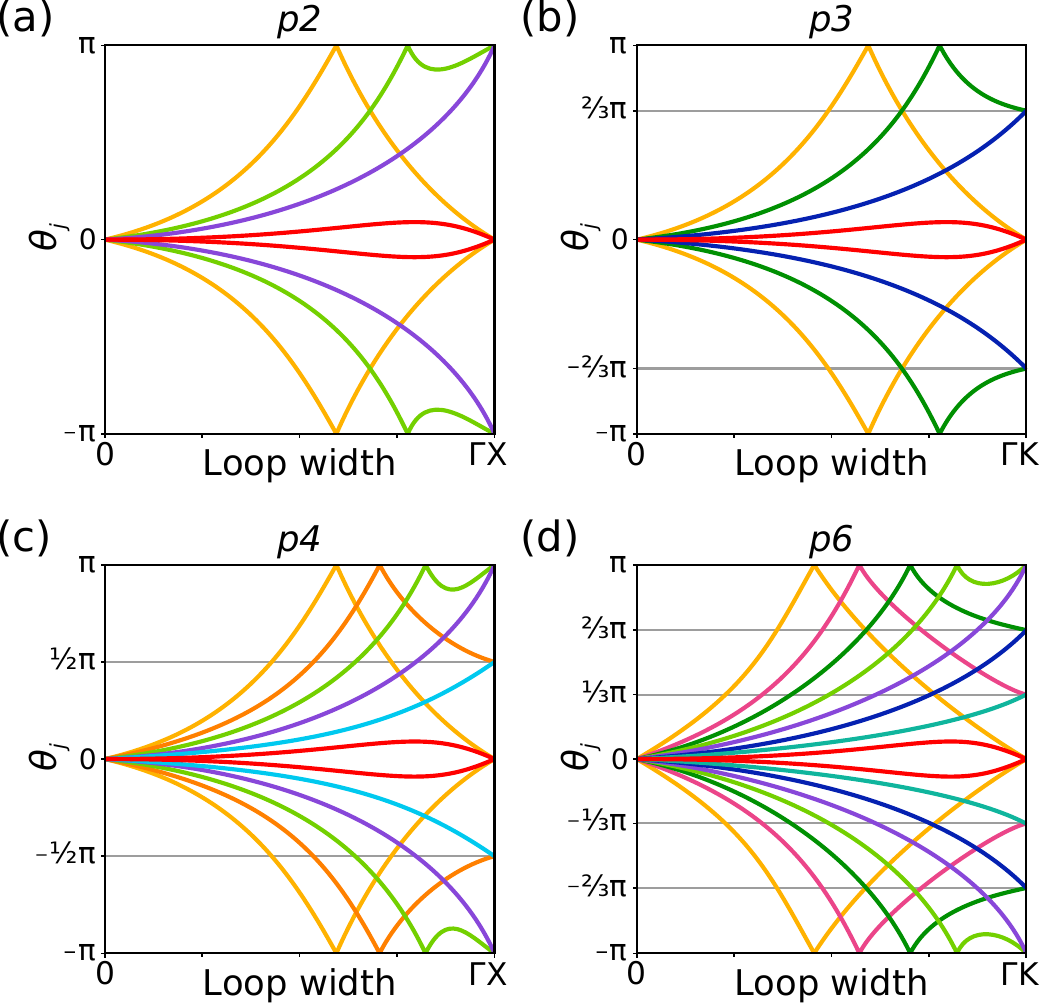}
    \caption{\label{fig:elementaryspectra} Sketch of all elementary Wilson loop spectra for systems with TRS and $n$-fold rotational symmetry, determined by concentric Wilson loops that expand from $\Gamma$ to covering $1/n^{\rm th}$ of the BZ. All pure rotation wallpaper groups are represented here, with $n$ equal to (a) $2$, (b) $3$, (c) $4$ and (d) $6$. The horizontal grey lines indicate allowed endpoints of the spectra, which correspond to $2\pi/n$.}
\end{figure}

The blue spectra in Fig.~\ref{fig:morebands} demonstrate that a concentric WLS in $p3$ that winds to $10\pi/3$ and one that winds to $2\pi/3$ both carry topological indices $(1,0)$. In the same way, a $(0,0)$ spectrum that does not wind is topologically equivalent to one that winds to $4\pi$. More generally, the topological invariants extracted from concentric WLS are cyclic, and whenever a spectrum contains multiple $\pi$-crossings, these crossings can pairwise annihilate upon the addition of a trivial Kramers pair. We shall call the spectra that remain unchanged upon hybridisation with a trivial WLS ``elementary spectra''. In systems with two, three, four, or six-fold rotational symmetry, one finds 4, 4, 6, and 8 elementary spectra, respectively, as shown in Fig.~\ref{fig:elementaryspectra}. 

One should note that although the elementary WLS always correspond to a unique set of $w_{\rm FKM}$ and $w_\pi$ invariants, the opposite is not true. For example, in $p6$ there are two elementary spectra for each combination of the two invariants. Each of these pairs of elementary WLS can be transformed into one another upon hybridization with a specific non-trivial additional Kramers pair. In $p4$ the same doubling of spectra corresponding to a set of invariants occurs for a subset of the elementary WLS.

To complete the evaluation of all the topological invariants for a TRS system with rotational symmetry, the WLS in some cases needs to be augmented by line invariants. In $p2$, the $\mathbb{Z}_2^4$ classification is determined by two independent LBO line invariants, the FKM invariant, and the new $w_\pi$ invariant. The latter two can be evaluated directly from the concentric WLS, where the four elementary spectra describe two independent $\mathbb{Z}_2$ invariants, while the line invariants are the eigenvalues of two additional Wilson loops along $\Gamma \rm X \Gamma$ and $\Gamma \rm Y \Gamma$. In $p3$, there are no line invariants, so the full classification is $\mathbb{Z}_2^2$ which is fully determined by $w_{\rm FKM}$ and $w_\pi$ extracted from the concentric WLS. (Note that this corrects a statement in \cite{J3_2019}: although the $U(1)$ vortices discussed there cannot be moved away from $K$, they can be smeared in a $C_3$ and TRS invariant fashion, leaving two rather than three $\mathbb{Z}_2$ invariants.) In $p4$, one expects a $\mathbb{Z}_2^3$ classification~\cite{J3_2019}, where only one independent LBO line invariant remains, which can be combined with the $w_{\rm FKM}$ and $w_\pi$ extracted from the six elementary WLS. Finally, in $p6$ we expect a classification of $\mathbb{Z}_2^3$~\cite{J3_2019}, where the third invariant is the LBO invariant along the $C_2$ and time-reversal invariant line $\Gamma \rm M \Gamma$.

\section{A second example in $p2$}\label{app:p2}

\begin{figure}[t]
    \centering
    \includegraphics[width=0.6\linewidth]{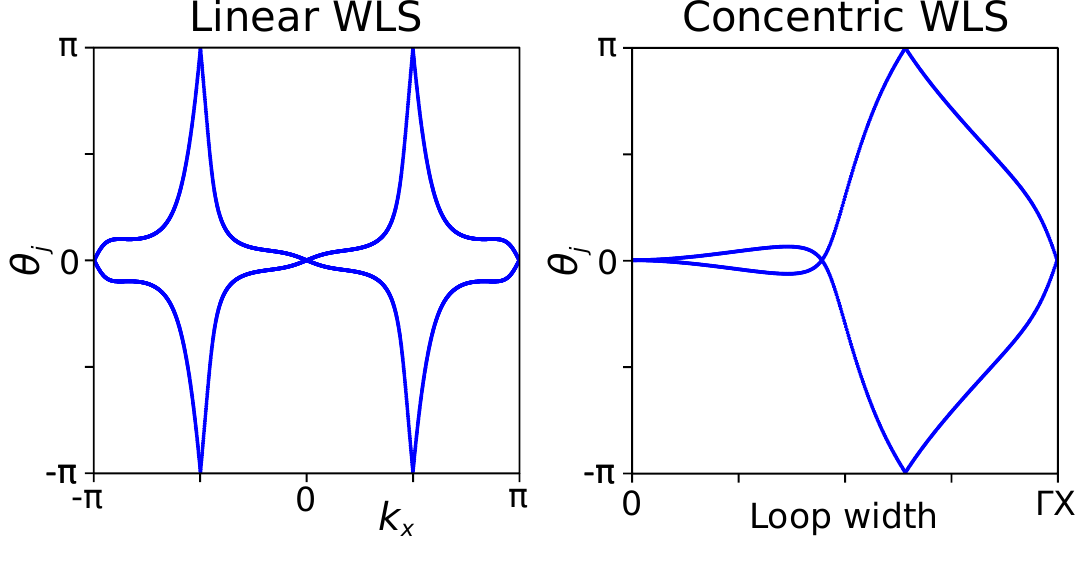}
    \caption{\label{fig:kmo-c2} A comparison of the linear and concentric WLS of the two-fold rotationally symmetric model considered in Ref.~\cite{Kooi2019}. The concentric WLS indicates the FKM and new topological invariants, which along with two independent LBO invariants completes the classification: $(w_{\rm FKM}, w_\pi, w_{\rm{LBO}_x}, w_{\rm{LBO}_y}) = (0,1,0,0)$.}
\end{figure}

In the main text, we have demonstrated the evaluation of our new invariant in a class AII system with two-fold rotational symmetry (wallpaper group $p2$). We demonstrate a second example of this in Fig.~\ref{fig:kmo-c2}, for the $C_2$-symmetric Hamiltonian used in Ref.~\cite{Kooi2019}, where we compare the linear WLS to a concentric WLS over half the Brillouin zone, and it is clear our new invariant is non-trivial ($w_\pi=1$). While this model was previously described to host ``fragile'' topology~\cite{Kooi2019}, we note that our new invariant is stable upon the addition of trivial Kramers pair. 

\section{Ribbon bandstructures of the extended Haldane model}\label{app:ribbon}

\begin{figure}[t]
    \centering
    \includegraphics[width=\linewidth]{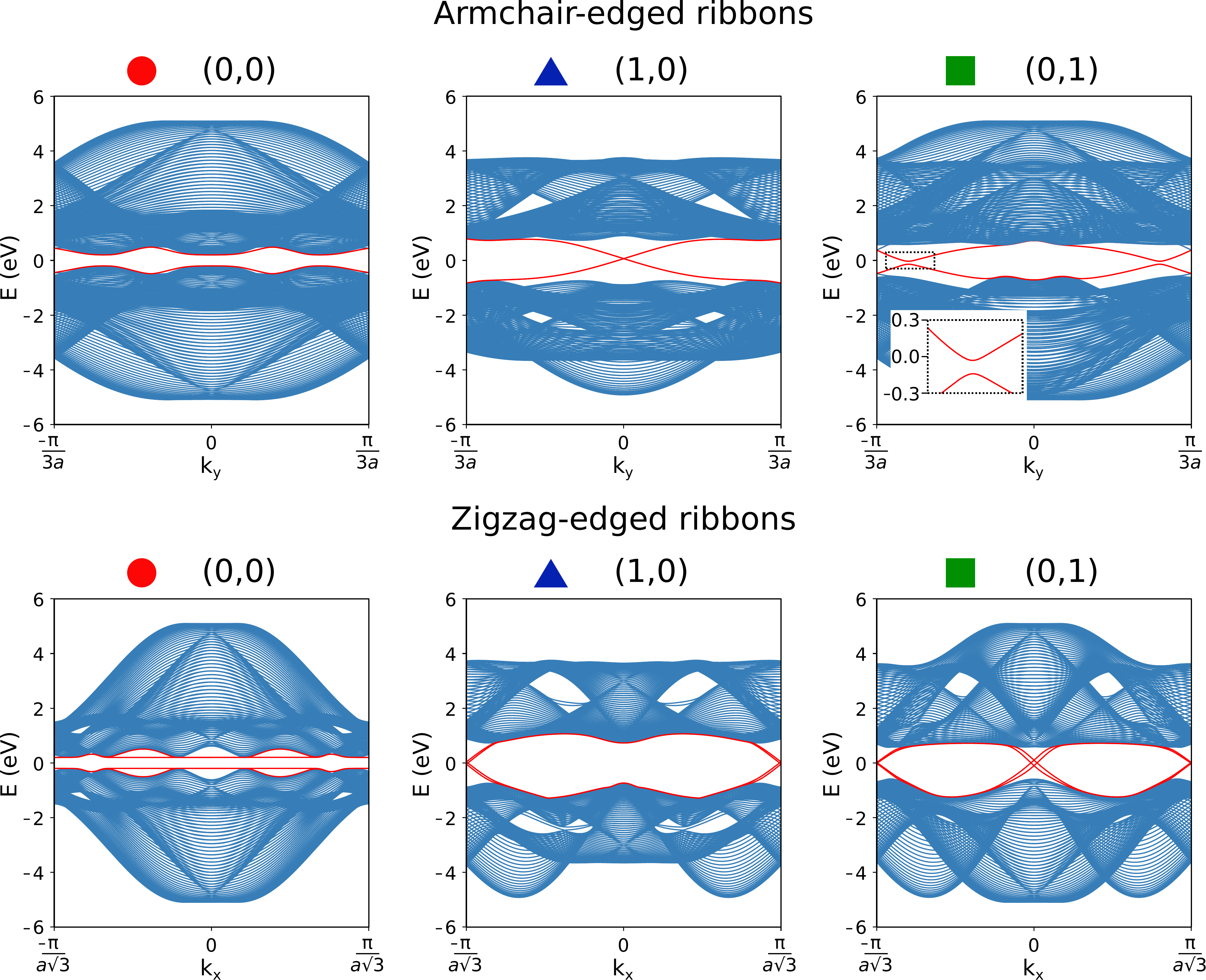}
    \caption{\label{fig:ribbons} Bandstructures for ribbons with armchair or zigzag edges, of the extended Haldane model with parameter values as indicated by the symbols in Fig. \ref{fig:p3} in the main text. The different parameter choices correspond to three different topological phases, with the indicated topological indices. Highlighted in red are the four states closest to the Fermi level.}
\end{figure}

To determine whether the three topological phases of the extended Haldane model (as described in the main text) host edge states, we consider two types of ribbon configurations. The honeycomb lattice structure of the system allows for two different (clean) terminations, commonly known as armchair or zigzag edges, which are constructed by breaking translational symmetry along the $x$ or $y$ directions, respectively. This distinction is important to make, because the two atoms in the unit cell (A and B) have different on-site energies ($\pm m$). While armchair edges host equal numbers of each of the A and B atoms on the edge, the two types of sites making up the zigzag edge will have different species of outermost atoms, breaking this sublattice symmetry. Figure \ref{fig:ribbons} shows numerically evaluated band structures for systems with both armchair and zigzag edges~\cite{Pybinding}, where we considered widths of 49 and 28 unit cells respectively, so that the ribbons have equal real-space width when the lattice geometry is taken into account.

In all three topological phases, the bulk system is an insulator, while edge states should show up in the ribbon geometry as in-gap states. In Fig. \ref{fig:ribbons}, the four states closest to $E=0$ are highlighted in red. In the $\mathbf{w}=(0,0)$ phase, these red states do not cross the gap and act like bulk states. For zigzag-edged ribbons, they become almost completely flat bands, which is likely due to charge localisation on the outermost atoms in the zigzag edges. This is made possible by the different on-site energies on A and B-type atoms in the unit cell. In contrast, the (1,0) phase contains edge states crossing the band gap, as expected for a system with non-trivial FKM invariant. Note that these edge states are two-fold degenerate in the armchair-edged ribbons, while this degeneracy is broken in the zigzag-edged ribbons. Finally, in the (0,1) phase, we find that the in-gap armchair edge states become gapped, while the zigzag edge states remain. In a finite-sized hexagon with armchair edges (Fig. \ref{fig:finite} in the main text), this phase hosts corner states whose energies can be tuned to fall within the spectral gap between the armchair edge states. 

An interesting area to further explore would be the protection of edge states against gapping in the various ribbon geometries. For the $(1,0)$ phase, the armchair edge states form a two-fold degenerate crossing, while the breaking of sublattice symmetry in the zigzag-edged ribbon splits the edge states into two non-degenerate crossings. Each such crossing contains two edge states, localized on opposite edges of the ribbon. These crossings are thus protected from opening a gap by the suppression of their overlap due to the large real-space distance between edges. States in different crossings on the other hand, are time-reversed partners, and cannot scatter into one another without breaking time-reversal symmetry. These same constraints remain in the degenerate armchair-edged ribbons, so that edge states in the $(1,0)$ phase are topologically protected.

In the $(0,1)$ phase, the number of zero energy crossings is doubled as compared to the $(1,0)$ phase. In this case, each side of the ribbon hosts two pairs of edge states connected by time-reversal symmetry. If two states on the same side of the ribbon that are not time-reversed partners can be brought together in $k_x$-space, they can be gapped. This happens in the armchair geometry. For the zigzag-edged ribbon in fig.~\ref{fig:ribbons}, states with the same real-space localization that are not time-reversed partners are still separated in momentum, and therefore not gapped. It may be possible to move the edge states crossings in $k_x$ space by tuning model parameters until they meet and open a gap by hybridising. Similarly, the addition of impurities that break the translational symmetry along the ribbon edge and allow scattering without conserving $k_x$ may also lead to a gap opening in the edge spectrum. In the presence of such a gap, it may be expected that corner charges could be stabilised also for finite hexagon-shaped systems with zigzag edges.

\end{document}